\documentclass[apj]{emulateapj}
\usepackage{color}

\newcommand{\eg}{e.g., }
\newcommand{\ie}{i.e., }
\newcommand{\Msun}{M_{\odot}}
\newcommand{\kms}{km~s$^{-1}$}

\newcommand{\Nifs}{$^{56}$Ni}

\def\gsim{\mathrel{\rlap{\lower 4pt \hbox{\hskip 1pt $\sim$}}\raise 1pt
\hbox {$>$}}}
\def\lsim{\mathrel{\rlap{\lower 4pt \hbox{\hskip 1pt $\sim$}}\raise 1pt
\hbox {$<$}}}

\def\ion#1#2{{\rm #1}~{\sc #2}}

\shorttitle{Spectropolarimetry of Extremely Luminous Type Ia SN 2009dc}
\shortauthors{Tanaka et al.}

\begin{document}

\title{
Spectropolarimetry of Extremely Luminous Type I\lowercase{a} 
Supernova 2009\lowercase{dc}: Nearly Spherical Explosion of 
Super-Chandrasekhar Mass White Dwarf \altaffilmark{1}}
\author{
Masaomi Tanaka\altaffilmark{2, 3}, 
Koji S. Kawabata\altaffilmark{4},
Masayuki Yamanaka\altaffilmark{5,4},
Keiichi Maeda\altaffilmark{3},
Takashi Hattori\altaffilmark{6},
Kentaro Aoki\altaffilmark{6},
Ken'ichi Nomoto\altaffilmark{3},
Masanori Iye\altaffilmark{7},
Toshiyuki Sasaki\altaffilmark{6},
Paolo A. Mazzali\altaffilmark{8,9}, and
Elena Pian\altaffilmark{10,11}
}

\altaffiltext{1}{Based on data collected at Subaru Telescope, 
which is operated by the National Astronomical Observatory of Japan.}
\altaffiltext{2}{Department of Astronomy, Graduate School of Science, University of Tokyo, Bunkyo-ku, Tokyo, Japan; mtanaka@astron.s.u-tokyo.ac.jp}
\altaffiltext{3}{Institute for the Physics and Mathematics of the Universe, University of Tokyo, Kashiwa, Japan}
\altaffiltext{4}{Hiroshima Astrophysical Science Center, Hiroshima University, Higashi-Hiroshima, Hiroshima, Japan}
\altaffiltext{5}{Department of Physical Science, Hiroshima University, Higashi-Hiroshima, Hiroshima, Japan}
\altaffiltext{6}{Subaru Telescope, National Astronomical Observatory of Japan, Hilo, HI}
\altaffiltext{7}{Optical and Infrared Astronomy Division, National Astronomical Observatory, Mitaka, Tokyo, Japan}
\altaffiltext{8}{Max-Planck Institut f\"ur Astrophysik, Karl-Schwarzschild-Strasse 2 D-85748 Garching bei M\"unchen, Germany}
\altaffiltext{9}{Istituto Naz. di Astrofisica-Oss. Astron., vicolo dell'Osservatorio, 5, 35122 Padova, Italy}
\altaffiltext{10}{Istituto Naz. di Astrofisica-Oss. Astron., Via Tiepolo, 11, 34131 Trieste, Italy}
\altaffiltext{11}{European Southern Observatory, Karl-Schwarzschild-Strasse 2, D-85748, Garching, Germany}

\begin{abstract}
We present the first spectropolarimetric observations of
a candidate of super-Chandrasekhar mass Type Ia supernova (SN):
SN 2009dc.
The observations were performed at 5.6 and 89.5 days after 
the $B$-band maximum. 
The data taken at the later epoch are used to determine 
the interstellar polarization.
Continuum polarization is found to be small ($<0.3 \%$),
indicating that the explosion is nearly spherically symmetric.
This fact suggests that a very aspherical explosion 
is not a likely scenario for SN 2009dc.
Polarization at the \ion{Si}{ii} and \ion{Ca}{ii} lines clearly
shows a loop in the $Q$-$U$ plane, indicating a non-axisymmetric, 
clumpy distribution of intermediate-mass elements.
The degree of line polarization at the Si and Ca line is moderate
($0.5 \% \pm 0.1 \%$ and $0.7 \% \pm 0.1 \%$, respectively),
but it is higher than expected from the trend of other Type Ia SNe.
This may suggest that there are thick enough, clumpy Si-rich layers
above the thick \Nifs-rich layers ($\gsim 1.2 \Msun$).
The observed spectropolarimetric properties, 
combined with the photometric and spectroscopic 
properties, suggest that 
the progenitor of SN 2009dc has a super-Chandrasekhar mass,
and that the explosion geometry is globally spherically symmetric,
with clumpy distribution of intermediate-mass elements. 
\end{abstract}


\section{Introduction}
\label{sec:introduction}
Type Ia supernova (SN) is a thermonuclear explosion of a C+O white dwarf
in a close binary \citep[see,][for a review]{hillebrandt00,nomoto94Ia}.
When the mass of the white dwarf becomes close to the 
Chandrasekhar mass ($\simeq 1.38 \Msun$), the explosion is triggered.
Thus, the ejecta mass of Type Ia SNe is always similar
\citep{mazzali07Ia}.
Accordingly, the observed properties of Type Ia SNe 
are more homogeneous than those of core-collapse SNe, which have
various ejecta masses.

Thanks to this homogeneity, Type Ia SNe are used as 
a cosmological distance indicator.
Although it is known that the maximum luminosity has a diversity,
this can be corrected using the shape of the light curve:
the brighter SNe decline more slowly \citep[\eg][]{phillips93,phillips99}.
Using Type Ia SNe as a distance indicator, 
the accelerating nature of the Universe has been revealed
\citep[\eg][]{riess98,perlmutter99}.

Recently, the discovery of a class of extremely luminous 
Type Ia SNe has casted the question on the progenitor scenario.
\citet{howell06} reported that SN 2003fg had a maximum brightness
$M_V = -19.9$ mag. 
To power such an extreme luminosity, 
$\sim 1.3 \Msun$ of \Nifs\ is required.
Despite the large \Nifs\ production, the expansion velocity 
of SN 2003fg ($\sim 8,000$ \kms\ near the maximum brightness) 
is slower than normal Type Ia SNe ($\sim 10,000$ \kms).
From these facts, they suggested that the progenitor of SN 2003fg
has a super-Chandrasekhar mass ($\sim 2 \Msun$).

Another candidate of super-Chandrasekhar mass Type Ia SN 
is SN 2006gz \citep{hicken07}.
The maximum brightness is $M_V \sim -19.74$ mag
when the host extinction $A_V = 0.56$ mag is assumed.
SN 2007if also shows spectral similarities to SN 2003fg
\citep[see][]{yuan07,akerlof07,rau07,scalzo10}.

This class of objects commonly shows clear \ion{C}{ii} lines in the 
optical spectrum.
The C lines are rarely seen in normal Type Ia SNe \citep{marion06,tanaka08Ia}.
The presence of C also seems to support 
a super-Chandrasekhar mass progenitor
since it suggests that there is a lot of unburned elements 
above the thick \Nifs-rich layer.
The C-rich ejecta might also explain the faintness of SN 2006gz
at late phases \citep{maeda09}.

However, there is a possibility that this class of objects is an 
aspherical explosion of a Chandrasekhar mass white dwarf.
\citet{hillebrandt07} discussed the difficulty in synthesizing 
much \Nifs\ in a differentially rotating white dwarf, 
and suggested an aspherical, lopsided explosion of a Chandrasekhar 
mass white dwarf.
\citet{maedaiwamoto09} argued that a combination of 
the low velocity and the moderate light curve decline rate of SN 2003fg is
not compatible with super-Chandrasekhar mass explosion, 
although the observed properties of SN 2006gz are consistent with
a spherical explosion of a super-Chandrasekhar mass progenitor.
They also suggested possible asphericity, if those two overluminous
SNe are to be explained in a unified scenario.
Asphericity must be one of the keys to understanding 
the nature of the extremely luminous Type Ia SNe.

Asphericity of extragalactic SNe can be studied by 
polarization measurements \citep[see][for a review]{wang08}.
Polarization is generated by the electron scattering in the SN ejecta.
When a photon is scattered by an electron with nearly $90^{\circ}$, 
unpolarized light gains large linear polarization, whose 
vector direction is orthogonal to the scattering plane.
Since extragalactic SNe are point sources, 
the polarization vectors are cancelled out in spherically symmetric ejecta.
Thus, if polarization is detected, it undoubtedly means asymmetry
of the SN ejecta \citep{shapiro82,mccall84,hoeflich91}.
The larger degree of polarization means larger deviation from
spherical symmetry.
Past observations show that the continuum polarization of 
normal Type Ia SNe is small, being $\lsim 0.3 \%$ 
\citep{wang97,wang0301el,wang06,leonard05,chornock08}.
Thus, the explosion of normal Type Ia SNe is almost spherical.

In addition, polarization at the line has information
on element distribution.
Scattering by lines tends to depolarize lights
\citep{jeffery89,kasen03}.
If the line scattering elements are distributed aspherically, 
the cancellation of polarization vectors becomes incomplete
at the absorption minimum of the P Cygni profile.
As a result, enhancement of the polarization degree is detected 
at the absorption.
In other words, detection of the line polarization directly means 
that the distribution of the elements is not uniform
in front of the photosphere.

Although spectropolarimetry is a powerful tool,
it was not performed for the three candidates of 
super-Chandrasekhar mass Type Ia SN so far.
In this paper, we present the first spectropolarimetric 
observation of a candidate of super-Chandrasekhar
mass Type Ia SN: SN 2009dc.

SN 2009dc was discovered by \citet{puckett09} in S0 galaxy UGC 10064.
\citet{harutyunyan09} performed the first spectroscopy and reported
the slow expansion velocity and the presence of the \ion{C}{ii} line.
\citet{marion09} performed infrared spectroscopy and 
reported non-detection of \ion{C}{i} lines.

\citet{yamanaka09} showed that 
SN 2009dc is an extremely luminous event.
If the distance modulus $\mu=34.88$ mag 
and Galactic extinction $A_V=0.22$ mag \citep{schlegel98}
are assumed, 
the peak absolute magnitude is $M_V=-19.9$ mag 
(even assuming null extinction in the host galaxy).
This is twice as bright as the average luminosity of Type Ia SNe.
The required mass of \Nifs\ is estimated to be $\gsim 1.2 \Msun$.
Thus, SN 2009dc belongs to the same class as SNe 2003fg and 2006gz:
the fourth candidate of super-Chandrasekhar mass Type Ia SN.
For the properties of SN 2009dc, see also a recent paper by \citet{silverman10}.

In this paper, we report the spectropolarimetric observations of SN 2009dc 
and study the explosion geometry and the progenitor of this SN.
In Section \ref{sec:obs}, we describe our observation and data
reduction.
In Section \ref{sec:results}, we show the results of the spectropolarimetry.
An estimate for interstellar polarization is also given.
We summarize the intrinsic properties of continuum and line polarization
of SN 2009dc in Section \ref{sec:intrinsic}.
Our spectropolarimetric observations suggest that
SN 2009dc is a nearly spherical explosion of 
a super-Chandrasekhar mass white dwarf.

\section{Observations and Data Reduction}
\label{sec:obs}

Spectropolarimetric observations of SN 2009dc were performed 
with the 8.2m Subaru telescope equipped
with Faint Object Camera and Spectrograph \citep[FOCAS,][]{kashikawa02}
on UT 2009 May 1.5 (MJD=54952.5) and July 24.4 (MJD=55036.4).
These epochs correspond to $t=+5.6$ and $+89.5$ days from the $B$ band
maximum (MJD=54946.9, \citealt{yamanaka09}).
Hereafter, $t$ denotes the days after the $B$-band maximum 
in the observer's frame.

For the observation at $t=+5.6$ days, 
we used an offset slit of $0.8''$ width,
a 300 lines mm$^{-1}$ grism, and the Y47 filter.
This configuration gives a rather wide wavelength coverage, 
4400-9000 \AA.
At $t=+89.5$ days, given the expected small polarization level,
we used a center slit of  $0.8''$ width 
(with the same grism without any order-sorting filter)
since the instrumental polarization of FOCAS is negligible
($\lsim 0.1 \%$) when the center slit is used \citep{tanaka0807gr}.
This configuration gives a wavelength coverage of 4000-7000 \AA.
The wavelength resolution is $\lambda/\Delta \lambda \sim 650$
in both settings.

For the measurement of linear polarization, 
FOCAS is equipped with a rotating superachromatic half-wave plate 
and a crystal quartz Wollaston prism.
An incident beam is split to 
ordinary and extraordinary lights by the Wollaston prism,
and both rays are recorded on the CCD simultaneously.
One set of observations consists of
the integration with 
$0^{\circ}, \ 45.^{\circ},\ 22.5^{\circ}$, and $67.5^{\circ}$ 
positions of the half-wave plate.
From this one set of exposures,
Stokes parameters $Q$ and $U$ can be derived
as described by \citet{tinbergen96}.

For the observations at $t=+5.6$ days, 
we performed three sets of four integrations
with the exposure time of 2400 s for each set 
(600 s $\times$ 4).
Also, one set was taken 
with the exposure time of 3600 s (900 s $\times$ 4).
At $t=+89.5$ days, 
we took three sets of four integrations with the exposure time
of 3200 s for each set (800 s $\times$ 4).
The total exposure time is 10800 s and 9600 s at $t=+5.6$ and $+89.5$
days, respectively.

We derived Stokes parameters for 
each set of observations independently.
And then, these were combined, rejecting the apparent noises
(by rejecting $> 3 \sigma$ points from the median of 
30-50 \AA\ window, and by further visual inspection).
When the offset slit is used ($t=+5.6$ days), 
the instrumental polarization of 
$\sim 0-0.4 \%$ is corrected by observing the
unpolarized stars BD+32$^{\circ}$3739 and BD+28$^{\circ}$4211 
\citep{schmidt92pol}.
The reference axis of the position angle was 
calibrated by the observation of 
the strongly polarized star HD 155197 \citep{turnshek90}.
The wavelength dependence of the optical axis of the half-wave plate
was corrected using the dome flat taken through 
a fully-polarizing filter.

The total flux was calibrated using the observation 
of the spectrophotometric standard star BD+28$^{\circ}$4211 \citep{oke90}
at both epochs.
Telluric absorption lines were also removed using 
the spectrum of this standard star.

Throughout this paper, we define Stokes parameters as
a fraction of the total flux: $Q \equiv \hat{Q}/I$ and 
$U \equiv \hat{U}/I$, where $\hat{Q}$ and $\hat{U}$ can be 
expressed by $\hat{Q} = I_{0} - I_{90}$ and
$\hat{U} = I_{45} - I_{135}$, respectively. 
Here $I_{\theta}$ is the intensity 
measured through the ideal polarization filter with an 
angle $\theta$ \citep[see \eg][]{landi02}.
The angle is measured from north to east.
From Stokes $Q$ and $U$,
the position angle is obtained by $\theta \equiv 0.5 {\rm atan}(U/Q)$.
The degree of total polarization is $P \equiv \sqrt{Q^2 + U^2}$.
Because of this definition, total polarization is biased 
in a positive direction. This bias is
corrected using the results of \citet{patat06} when $P$ is shown.

\begin{figure}
\begin{center}
\includegraphics[scale=0.44]{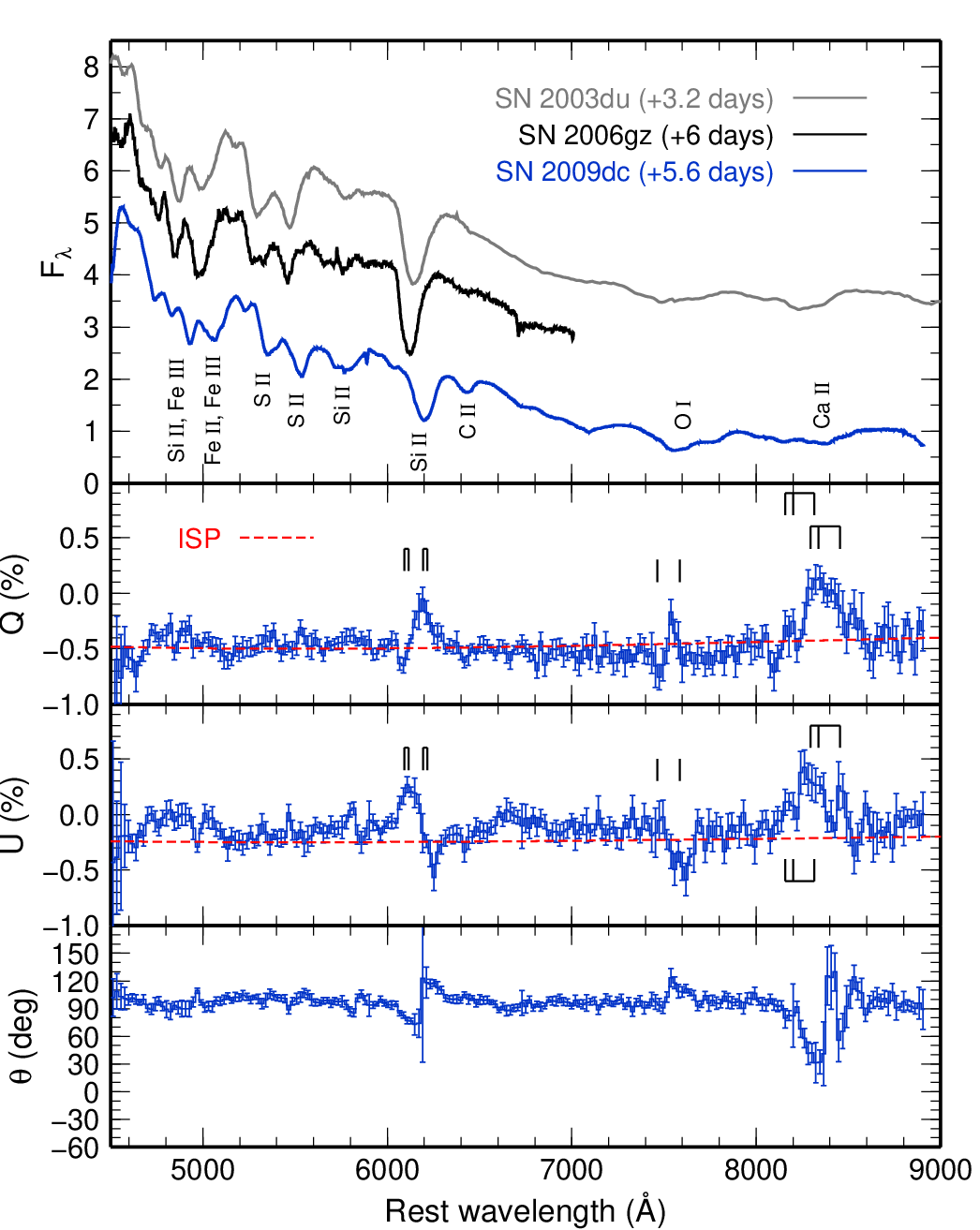}
\caption{
Total flux and polarization spectrum of SN 2009dc at $t=+5.6$ days
(blue lines).
In polarization spectrum, ISP is {\it not} corrected for.
Polarization data are binned into 20 \AA.
In the top panel, the total flux of SN 2009dc 
(in $10^{-15}\ {\rm erg\ s^{-1}\ cm^{-2}\ \AA^{-1}}$) is compared with 
the normal Type Ia SN 2003du 
\citep[scaled flux, shifted by 3.0,][]{stanishev07} and 
the overluminous Type Ia SN 2006gz 
\citep[scaled flux, shifted by 1.5,][]{hicken07}.
The red dashed lines show the estimated ISP (Section \ref{sec:ISP}).
Vertical lines at the \ion{Si}{ii} ($\lambda$6347, 6371),
\ion{O}{i} $\lambda$7774, 
and \ion{Ca}{ii} ($\lambda$8498, 8542, 8662) lines show
the 7,200 \kms\ and 12,000 \kms\ positions.
\label{fig:pol}}
\end{center}
\end{figure}

\begin{figure}
\begin{center}
\includegraphics[scale=0.44]{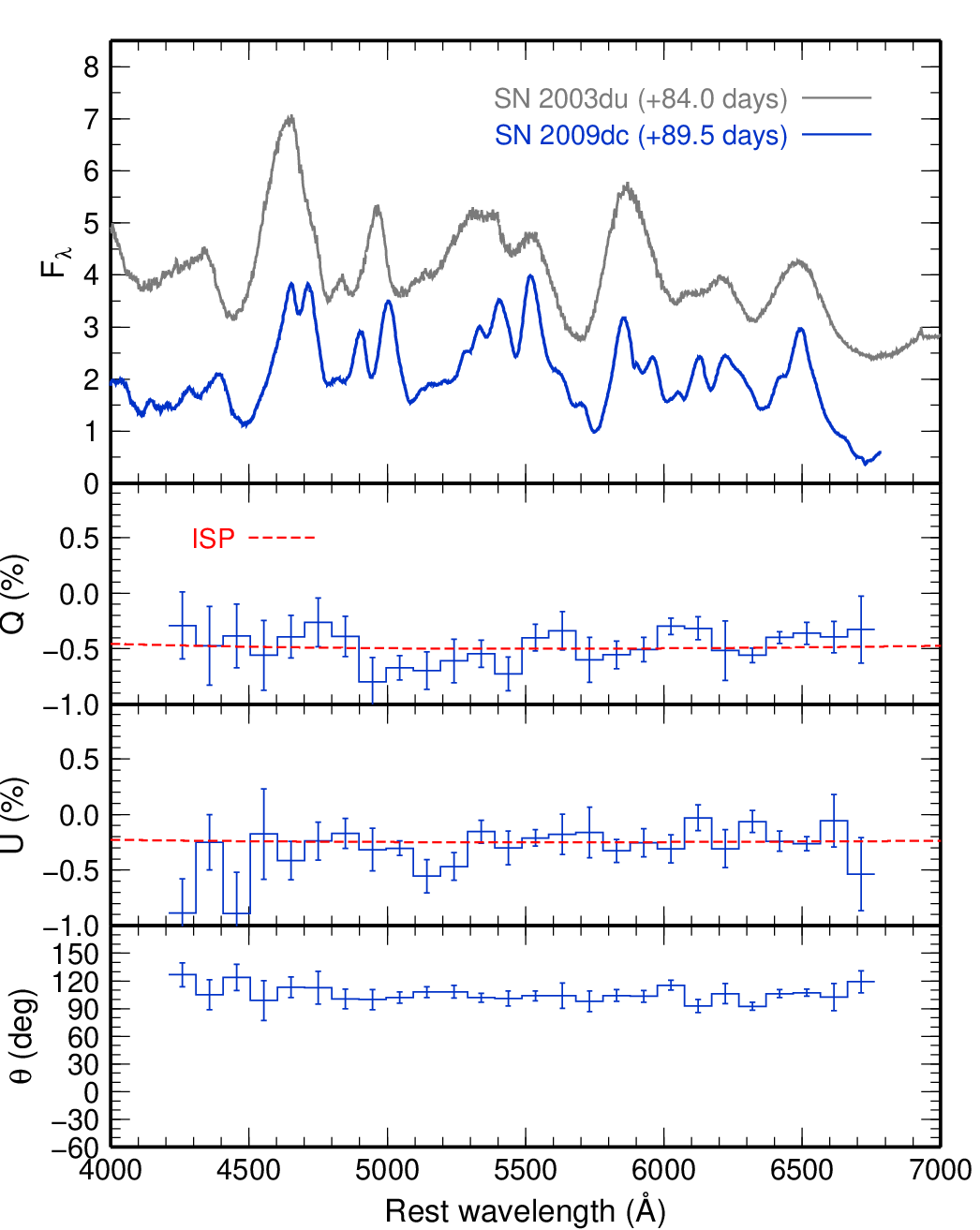}
\caption{
Total flux and polarization spectrum of SN 2009dc at $t=+89.5$ days
(blue lines).
In polarization spectrum, ISP is {\it not} corrected for.
Polarization data are binned into 100 \AA.
In the top panel, the total flux of SN 2009dc 
(in $10^{-16}\ {\rm erg\ s^{-1}\ cm^{-2}\ \AA^{-1}}$) is compared with 
the normal Type Ia SN 2003du 
\citep[scaled flux, shifted by 2.0,][]{stanishev07}.
The red dashed lines show the estimated ISP (Section \ref{sec:ISP}).
\label{fig:pol_p90}}
\end{center}
\end{figure}

\section{Results}
\label{sec:results}

\subsection{Spectroscopic Properties}
\label{sec:spec}

The {\it top} panel of Figure \ref{fig:pol} shows 
the total flux spectrum of SN 2009dc at $t=+5.6$ days (blue) compared with
the spectra of normal SN 2003du \citep[gray,][]{stanishev07} 
and overluminous SN 2006gz \citep[black,][]{hicken07}.
SN 2003du is a well-studied normal Type Ia SN 
\citep[$z=0.00638$,][]{gerardy04,anupama0503du,stanishev07}.
SN 2006gz is an overluminous Type Ia SN, which is a
candidate for super-Chandrasekhar mass SN 
\citep[$z=0.0236$,][]{hicken07}.
Compared with normal Type Ia SNe, 
the spectrum of SN 2009dc is unique in the following points:
(1) very low line velocities and (2) presence of the carbon line
(see also \citealt{yamanaka09}).

The velocity of the \ion{Si}{ii} $\lambda$6355 line is 
$v=7,200$ \kms, lower than that of SN 2003du at a similar epoch
(10,200 \kms\ at $+6.2$ days) and SN 2006gz (11,200 \kms\ at $+6$ days).
The Si line velocity of SN 2009dc is close to that of SN 2003fg
\citep[8000 \kms\ at $+2$ days,][]{howell06}.

The \ion{C}{ii} $\lambda$6578 line is clearly present in the spectrum
of SN 2009dc at $+5.6$ days.
The velocity is $v=6,700$ \kms, which is also very low for a Type Ia SN.
The C lines are rarely seen in Type Ia SNe \citep{marion06,tanaka08Ia}.
Even if they are seen, the detection is limited almost exclusively to 
the very early phases: the earliest detection was in SN 1990N at $-15$ days 
\citep{mazzali0190N}, and the detection 
is usually at $<-10$ days for other SNe \citep{thomas07}.
In fact, the \ion{C}{ii} line was also visible in SN 2006gz, but
disappeared at $-10$ days.
In SN 2003fg, the \ion{C}{ii} line was marginally present at $+2$ days.

Figure \ref{fig:pol_p90} shows the total flux spectrum of SN 2009dc 
at $t=+89.6$ days compared with SN 2003du at $t=84.0$ days
\citep{stanishev07}.
This epoch is a transition phase when the SN spectrum evolves 
from a photospheric spectrum to a nebular spectrum.
In the spectra of both SNe, 
continuum emission is nearly absent,
and emission features start to dominate the spectrum.
Line features of SN 2009dc are narrower than 
those of SN 2003du even at this epoch.

\begin{figure}
\begin{center}
\includegraphics[scale=0.8]{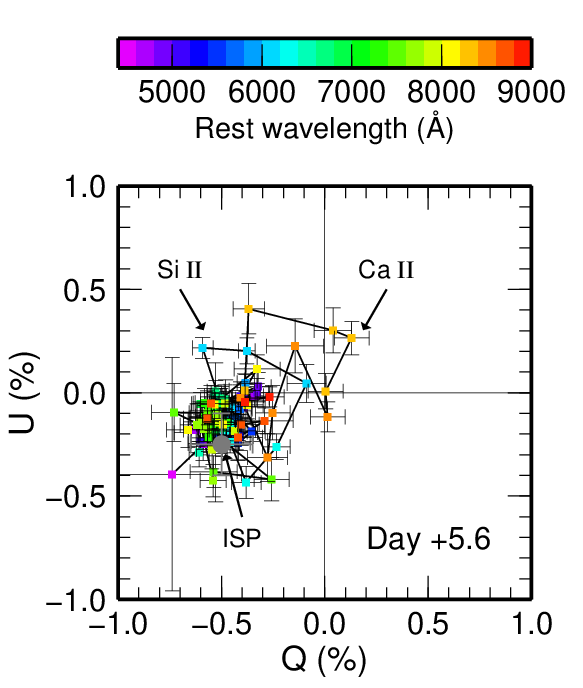}
\caption{
Polarization data at $t=+5.6$ days in the $Q$-$U$ plane.
Different colors show the wavelength according to the color scale bar.
ISP at 5500 \AA\ is marked with the gray point.
The data are binned into 40 \AA.
\label{fig:qu}}
\end{center}
\end{figure}

\subsection{Spectropolarimetric Properties}
\label{sec:specpol}

Observed Stokes $Q$ and $U$ at $t=+5.6$ days 
are shown in the middle panels of Figure \ref{fig:pol}.
Polarization data are binned into 20 \AA.
It clearly shows a variation in the polarization degree at
the \ion{Si}{ii} $\lambda$6355 and the \ion{Ca}{ii} IR triplet lines.
The change in the degree of polarization at these lines is $\sim$ 0.5\%
(see Section \ref{sec:line} for more details),
which is broadly consistent with the past measurements
for other Type Ia SNe
\citep{wang97,wang0301el,wang07,leonard05,chornock08}.
A smaller fluctuation is also seen at the \ion{O}{i} $\lambda$7774.

Except for these lines, the degree of polarization is
$Q \sim -0.5\%$ and $U \sim -0.2\%$.
No significant polarization variation is detected at the \ion{C}{ii} line
and in the wavelength region of $< 6000$ \AA, where
the \ion{S}{ii} lines and many \ion{Fe}{ii}-\ion{Fe}{iii} lines are present.

The polarization data at $t=+5.6$ days are plotted 
in the $Q$-$U$ plane in Figure \ref{fig:qu}.
This also shows that the majority of the data is confined 
around $Q \sim -0.5\%$ and $U \sim -0.2\%$. 
And, the polarization around the \ion{Si}{ii} and \ion{Ca}{ii} 
lines deviates from this point, making a loop in the $Q$-$U$ plane.
This is also inferred from large variation in $\theta$ at these 
lines (Figure \ref{fig:pol}, see Section \ref{sec:line} for more details).
According to the classification of spectropolarimetry types
by \citet{wang08}, SN 2009dc shows type N1 for the continuum 
(no significant dominant axis)
and type L for the \ion{Si}{ii} and \ion{Ca}{ii} lines 
(large polarization change at the line).

Stokes $Q$ and $U$ at the \ion{Ca}{ii} IR triplet seem to have 
substructures, \ie the main peak around 8300 \AA\ 
and a small peak at 8450 \AA. To identify these peaks, 
the position of three lines ($\lambda$8498, 8542, 8662) are marked
with the Doppler shifts of $v=7,200$ \kms\ (photospheric component) 
and $v=12,000$ \kms\ (high velocity component) in Figure \ref{fig:pol}.
It is found that the small peak at 8450 \AA\ originates 
from the photospheric component
of the reddest \ion{Ca}{ii} $\lambda$8662, while the main peak is 
a combination of the photospheric component of the other two lines
and the high velocity component of the three lines.
Similar substructures were also seen in the polarization spectrum
of SN 2004S at $+9$ days \citep{chornock08}, but 
polarization of the high velocity component is smaller than the photospheric
component in SN 2004S.

We also mark these two velocity components for the \ion{Si}{ii} line.
Stokes $U$ has a strong peak at a similarly 
high velocity ($v=12,000$ \kms) and polarization variation 
is seen up to $v\sim 15,000$ \kms.
Note that the total flux has the absorption minimum at $v=7,200$ \kms\
and does not show a strong blue wing.

Figure \ref{fig:pol_p90} shows the polarization spectrum 
at $t=+89.5$ days.
Given the low S/N ratio, the data are binned into 100 \AA.
The overall polarization level is similar to that at $t=+5.6$ days
($Q \sim -0.5\%$ and $U \sim -0.2\%$).
No significant variation (with $\gsim 0.5$ \% level) of polarization 
is detected at the lines.
Also, the polarization angle $\theta$ is nearly constant.
Spectropolarimetry type at this epoch is type N1.

\subsection{Interstellar Polarization}
\label{sec:ISP}

The spectropolarimetric data suffer from interstellar polarization (ISP)
both in our Galaxy and the host galaxy.
To discuss intrinsic properties of the SN polarization, 
ISP must be corrected.
In this section, we give an estimate of ISP.

There are several possible methods to estimate ISP.
When only single-epoch data are available, 
and if there is a dominant axis (a straight line in the $Q$-$U$ plane)
in the polarization data, we may assume global axisymmetry of 
the SN ejecta and take an ISP at either end of the line
in the $Q$-$U$ plane
\citep{howell01,wang01,wang0302ap,tanaka0807gr}.
Another method is assuming complete depolarization at 
strong emission lines \citep[\eg][]{kawabata02,leonard05,wang06,maund0705bf}. 
This method is justified because 
the emission part of P Cygni profile consists largely
of line-scattered, depolarized photons.
However, it must be carefully done because 
depolarization is not necessarily complete \citep{tanaka0905bf}.

To avoid the uncertainty caused by these methods, 
we use spectropolarimetric data taken
at $t=89.5$ days.
As shown in Figure \ref{fig:pol_p90}, 
the spectrum at this epoch does not have clear continuum light,
and electron scattering is not important.
In addition, the emission features, 
which are intrinsically unpolarized, dominate the spectrum.
Thus, the light from SN at this epoch is assumed to be 
intrinsically unpolarized, and the observed polarization 
can be considered as ISP origin.

For the wavelength dependence of ISP, we assume 
the following relation,
which is valid for the Milky way like dust \citep{serkowski75}:
\begin{equation}
P(\lambda)\ 
=\ P_{\rm max}\exp \left[ 
-K\ {\rm ln}^2 \left( \lambda_{\rm max}/ \lambda \right) \right].
\end{equation}
Here $\lambda_{\rm max}$ is the wavelength at the peak of the ISP, 
$P_{\rm max}$ is the degree of the ISP at $\lambda_{\rm max}$,
and $K$ is given as 
$K = 0.01 + 1.66\ \lambda_{\rm max}\ ({\rm \mu m})$ \citep{whittet92}.
We assume $\lambda_{\rm max}=5500$ \AA\ as a conventional choice.

The best simultaneous fit of the Stokes $Q$ and $U$ at $t=+89.5$ days 
can be obtained ($\chi^2/dof$=1.1) when 
$Q_{\rm ISP} = -0.50 \%$ and 
$U_{\rm ISP} = -0.25 \%$ at 5500 \AA, 
\ie $P_{\rm ISP, max} = 0.56 \%$, $\theta_{\rm ISP} = 103^{\circ}$.
This ISP is shown in red dashed lines in Figures \ref{fig:pol}
and \ref{fig:pol_p90} and gray circle in Figure \ref{fig:qu}.
Note that the uncertainty of the ISP is $\sim 0.15 \%$, 
which is a typical error of the Stokes $Q$ and $U$
($\chi^2/dof \lsim 2.0$ with this uncertainty range).

The degree of ISP is also constrained by the total extinction: 
$P /E(B-V) < 9 \%$ \citep{serkowski75}.
In the line of sight to SN 2009dc, the extinction by our Galaxy is
$E(B-V)_{\rm Gal}=0.071$ mag \citep{schlegel98}.
The extinction by the host galaxy is not certain.
In our spectrum, narrow \ion{Na}{i} D lines are detected, 
and their equivalent widths are EW= 0.5 and 1.0 \AA\ for our Galaxy
and the host galaxy, respectively
(these are slightly lower than the values reported by \citealt{marion09}).
If we assume a host extinction proportional to the EW,
the host extinction is $E(B-V)_{\rm host}=0.14$ mag.
Alternatively, if we use the relations by \citet{turatto03},
we obtain  $E(B-V)_{\rm host}=0.15$ mag or $0.47$ mag.
Thus, the upper limit for the ISP is $P < 0.63 \%$ in 
our Galaxy and $P<1.3 \%$ (tightest when $E(B-V)=0.14$ mag)
in the host galaxy.
The ISP estimated above is consistent with these upper limits.

\begin{figure}
\begin{center}
\includegraphics[scale=0.44]{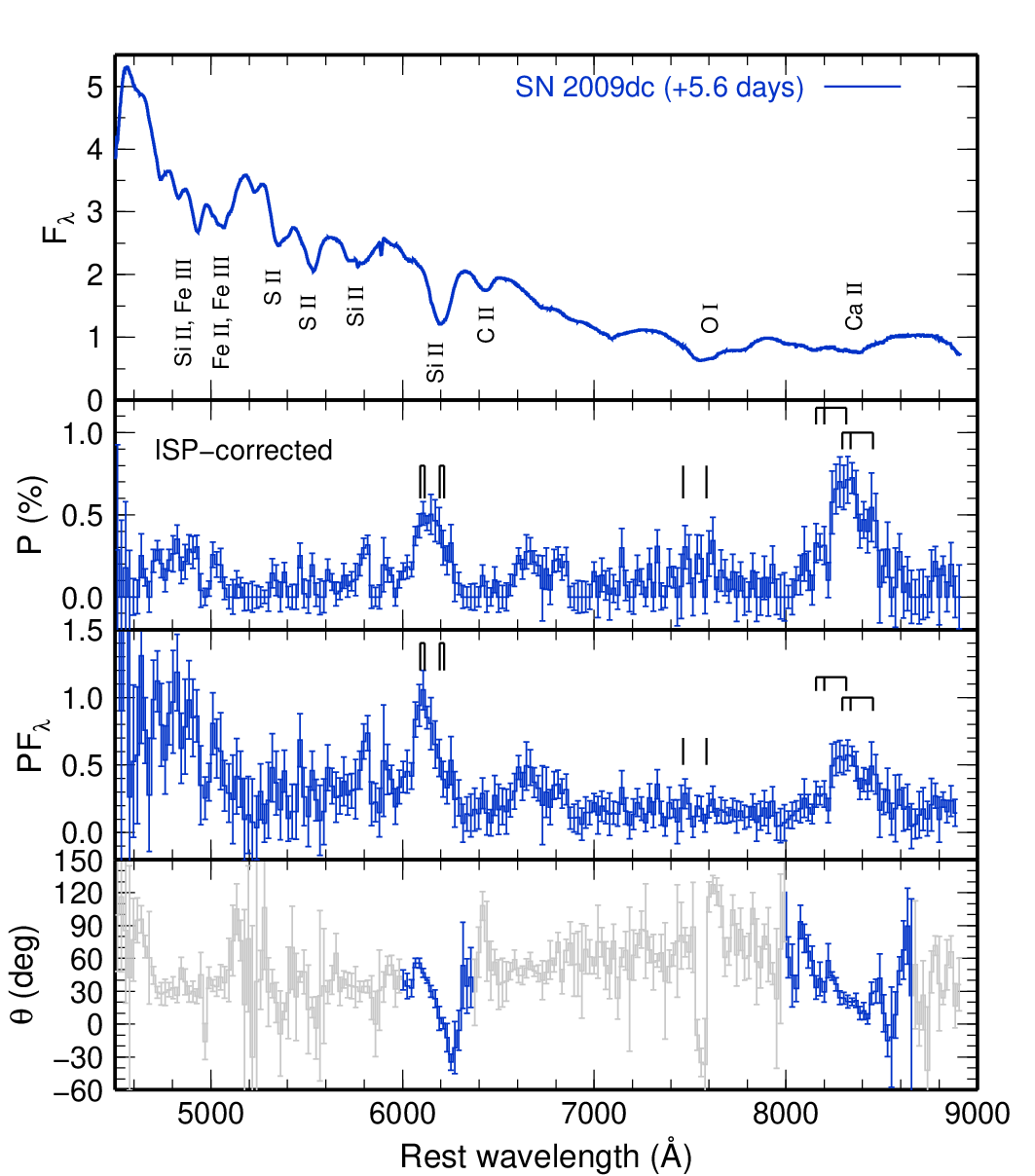}
\caption{
Total flux ($F_{\lambda}$ in $10^{-15} {\rm erg\ s^{-1}\ cm^{-2}\ \AA^{-1}}$), 
ISP-corrected, bias-corrected 
polarization $P$, polarized flux 
($PF_{\lambda}$ in $10^{-17} {\rm erg\ s^{-1}\ cm^{-2}\ \AA^{-1}}$), 
and position angle $\theta$ of SN 2009dc at $t=+5.6$ days.
In the plot of position angle,
the data around the Si and Ca lines are 
highlighted because, except for these parts,  
position angle is erratically scattered and not meaningful
due to the almost null polarization.
Vertical lines at the \ion{Si}{ii} ($\lambda$6347, 6371),
\ion{O}{i} $\lambda$7774, 
and \ion{Ca}{ii} ($\lambda$8498, 8542, 8662) lines show
the 7,200 \kms\ and 12,000 \kms\ positions.
\label{fig:ISP}}
\end{center}
\end{figure}

\section{Explosion Geometry and Progenitor of SN 2009\lowercase{dc}}
\label{sec:intrinsic}

\begin{figure*}
\begin{center}
\begin{tabular}{ccc}
\includegraphics[scale=0.55]{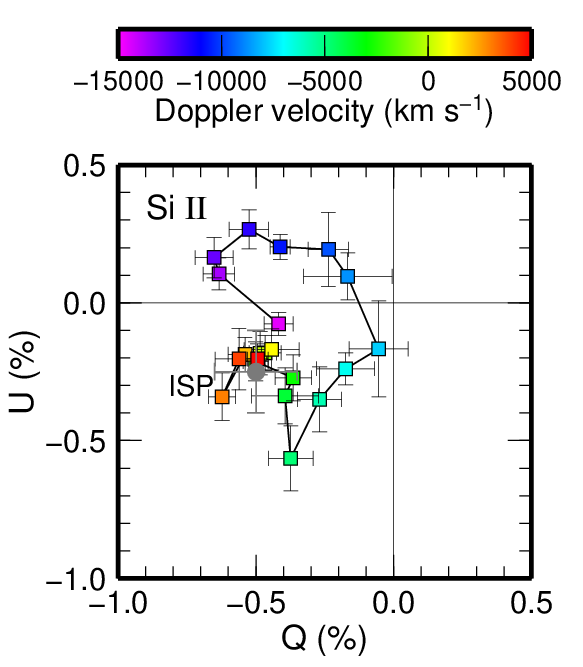}&
\includegraphics[scale=0.55]{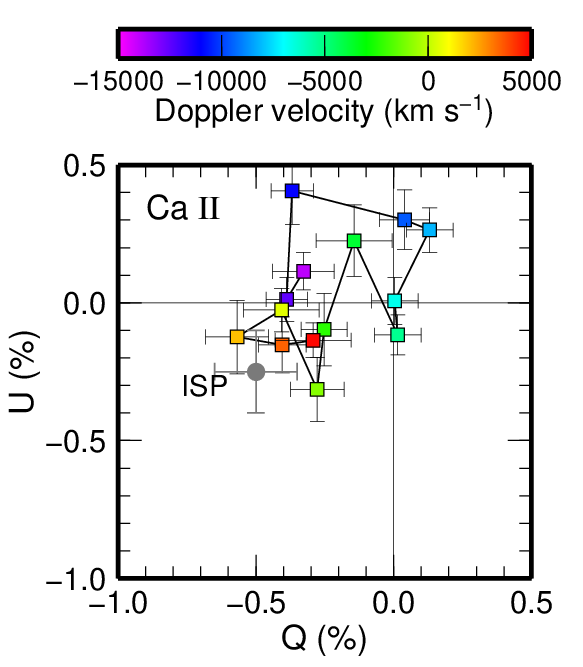}&
\includegraphics[scale=0.55]{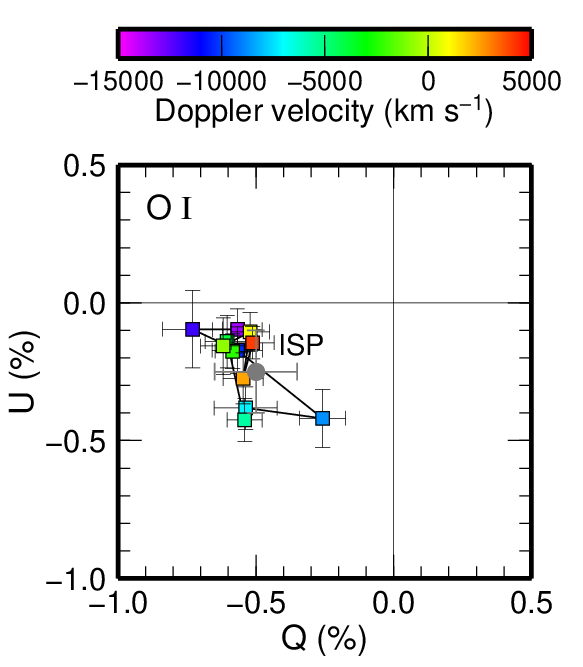}
\end{tabular}
\caption{
$Q$-$U$ diagram for the data around the \ion{Si}{ii} $\lambda$6355, 
\ion{Ca}{ii} IR triplet, and \ion{O}{i} $\lambda$7774
at $t=+5.6$ days.
Different color shows the Doppler velocity according to the 
color scale bar.
Note that the velocity is measured from 6355 \AA\ and 8567 \AA\
for the \ion{Si}{ii} ($\lambda$6347, 6371) 
and \ion{Ca}{ii} ($\lambda$8498, 8542, 8662) lines, respectively.
The data are binned into 20 \AA\ for the \ion{Si}{ii} line 
and 40 \AA\ for the \ion{Ca}{ii} and \ion{O}{i} lines.
\label{fig:lines}}
\end{center}
\end{figure*}

\subsection{Overall Asymmetry}
\label{sec:continuum}

Figure \ref{fig:ISP} shows the ISP-corrected, bias-corrected 
total polarization $P$ at $t=5.6$ days.
Assuming the ISP estimated with the later epoch data, 
the continuum polarization is found to be $< 0.3 \%$ 
measured at the line-free region at 6500-7300 \AA.
Although there is a fluctuation in polarization around 6600 \AA\
(this is less significant in the original Stokes $Q$ and $U$, 
Figure \ref{fig:pol}),
there is no corresponding line there.
Thus, we take this part as an upper limit of 
the continuum polarization.

The small degree of continuum polarization is 
consistent with other normal Type Ia SNe
\citep[\eg][]{wang97,wang0301el,wang06,leonard05,chornock08}.
The small degree of continuum polarization indicates
that the overall shape of the photosphere 
is nearly spherically symmetric.
If a pure electron-scattering ellipsoid is assumed 
and if it is viewed near 90 degrees,
the axis ratio of the photosphere deviates less than 10\% from unity
\citep{hoeflich91,wang97}.

\subsection{Element Distribution}
\label{sec:line}

Variation of polarization degree at the lines
is clearly detected 
at \ion{Si}{ii} $\lambda$6355 and the \ion{Ca}{ii} IR triplet, 
and \ion{O}{i} $\lambda$7774 in the data taken at $t=+5.6$ days
(Figure \ref{fig:pol}).
The line polarization means that the distribution of 
these elements are not spherically symmetric.
The degree of the total polarization is $0.5 \% \pm 0.1\%$ 
for the \ion{Si}{ii} line
and $0.7 \% \pm 0.1\%$ for the \ion{Ca}{ii} line
(ISP-corrected $P$, Figure \ref{fig:ISP}).
The polarization change at the \ion{O}{i} line is affected by
the choice of the ISP 
($P \lsim 0.3 \%$ in ISP-corrected $P$, Figure \ref{fig:ISP}), 
but fluctuation in the original Stokes $Q$ and $U$ is 
$\sim 0.2 \%$ (Figure \ref{fig:pol}).
The degree of the line polarization is similar to
that of normally luminous Type Ia SNe with normal expansion velocities.

If we look at the ISP-corrected polarization (Figure \ref{fig:ISP}), 
the smaller polarization change 
is also marginally seen at \ion{Si}{ii}-\ion{Fe}{iii} absorption 
at 4800\AA\ and \ion{Si}{ii} $\lambda$5789.
But these are not significant, 
and their significance does depend on the choice of the ISP.

Polarization at the \ion{Si}{ii}, \ion{Ca}{ii}, 
and \ion{O}{i} lines is plotted in the $Q$-$U$ plane
in Figure \ref{fig:lines}.
In these plots, the ISP is not corrected for, and 
the position of the ISP at 5500 \AA\ is shown by the gray circle.
Colors of the points represent the Doppler velocity measured from 
6355 \AA\ (\ion{Si}{ii}), 8567 \AA\ (\ion{Ca}{ii}), and 
7774 \AA\ (\ion{O}{i}).
As already shown in Section \ref{sec:specpol},
polarization change at the Si and Ca lines 
is detected not only at photospheric velocity ($v=7,200$ \kms),
but also at higher velocities ($v \gsim 12,000$ \kms).
This is more clearly seen in 
the polarized flux ($PF_{\lambda}$ in Figure \ref{fig:ISP}), 
which has a peak at high velocity ($v\sim 12,000$ \kms).

In the $Q$-$U$ plane (Figure \ref{fig:lines}), 
polarization at the \ion{Si}{ii} 
and \ion{Ca}{ii} lines shows a ``loop''.
It can also be inferred by the large variation in
the ISP-corrected position angle $\theta$ at the line 
(Figure \ref{fig:ISP}).
A similar structure in the $Q$-$U$ plane has been 
discovered in several Type Ia SNe, \eg SN 2001el \citep{wang0301el}
and SN 2004S \citep{chornock08}.
The presence of the loop means that 
the polarization degree and position angle do depend on the 
velocity, \ie the depth in the ejecta.
Thus, it suggests that the distribution of the element
is not axisymmetric.
If the distribution were axisymmetric, the data
would be distributed in a straight line in the $Q$-$U$ plane.

Interestingly, the loop of the \ion{Si}{ii} line 
and the \ion{Ca}{ii} line shares a similarity:
it goes counterclockwise for the larger blueshift
(higher velocity, see left and center panel of Figure \ref{fig:lines}).
This is in contrast to the behavior of the \ion{O}{i} line,
whose loop is not prominent (right panel of Figure \ref{fig:lines}). 
The marginal loop of the \ion{O}{i} line goes to a direction
different from that of the \ion{Si}{ii} and \ion{Ca}{ii} lines.

These facts suggest that the distributions of Si and Ca 
are correlated.
In addition, the distribution of these elements is
different from that of O, which may have a complementary distribution.
But all of these elements share similar velocity space.
This configuration can be understood if intermediate-mass elements
have a non-axisymmetric, large-scale clumpy distribution,
with some clumps sticking out to the O-rich layer \citep{wang06,wang08}.

Summarizing the spectropolarimetric properties,
the explosion geometry of SN 2009dc
seems to be similar to normally luminous Type Ia SNe
with normal expansion velocities.
The explosion is globally spherical with clumpy 
distribution of intermediate-mass elements.

\subsection{$P_{\rm Si~{\sc II}}$ - $\Delta m_{15}(B)$ Diagram}
\label{sec:Pdm15}

\begin{figure}
\begin{center}
\includegraphics[scale=0.8]{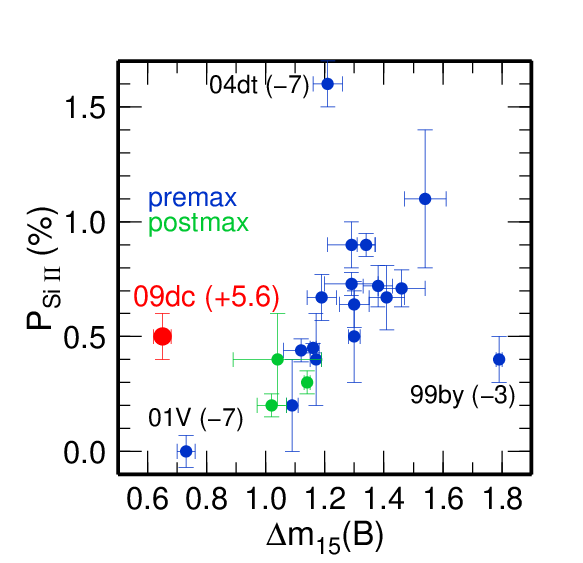}
\caption{
Polarization change at the \ion{Si}{ii} $\lambda$6355 line versus
the decline rate of the $B$ band light curve, $\Delta m_{15}(B)$.
The red point shows SN 2009dc.
The blue points show the data from \citet{wang07}.
The time evolution is {\it not} corrected for. 
They consist only of premaximum data (from $-10$ to $-2$ days).
The green points show the postmaximum data 
of SNe 1997dt ($+21$ days, \citealt{leonard05}),
2003du ($+18$ days, \citealt{leonard05}), 
and 2004S ($+9$ days, \citealt{chornock08}).
\label{fig:pdm15}}
\end{center}
\end{figure}

The clumpy structure of intermediate-mass elements 
discussed in Section \ref{sec:line} 
seems to be common in Type Ia SNe
\citep{wang06,wang08,chornock08}.
In this section, we discuss the property of line polarization of SN 2009dc 
comparing with other Type Ia SNe.

\citet{wang07} presented the largest spectropolarimetric sample
of Type Ia SNe. They showed a correlation between
the polarization degree of the \ion{Si}{ii} line and 
the decline parameter of the light curve $\Delta m_{15}(B)$, 
the difference in the $B$-band magnitude 
at maximum and at 15 days after the maximum
(brighter SNe decline more slowly, and thus, have
smaller $\Delta m_{15}(B)$, \eg \citealt{phillips93,phillips99}).
The data taken from \citet{wang07} are plotted in blue points
in Figure \ref{fig:pdm15}. 
The time evolution is not corrected for.
With two exceptions (SNe 2004dt and 1999by), they suggested 
that the brighter SNe (with smaller $\Delta m_{15}(B)$) show 
the smaller line polarization.

\citet{wang07} limited their sample to the data 
taken at premaximum phases.
From the literatures, we have added three spectropolarimetric examples
at postmaximum phases (green points in Figure \ref{fig:pdm15}):
SN 1997dt (+21 days, $P_{\rm Si {\sc II}} = 0.4 \% \pm 0.2\%$, 
\citealt{leonard05}),
SN 2003du (+18 days, $P_{\rm Si {\sc II}} = 0.2 \% \pm 0.05 \%$,
\citealt{leonard05}),
and SN 2004S 
(+4 days, $P_{\rm Si {\sc II}} = 0.3 \% \pm 0.05 \%$,
\citealt{chornock08}).
Their $\Delta m_{15}(B)$ is $1.04 \pm 0.15$ mag \citep{jha06}, 
$1.02 \pm 0.05$ mag \citep{anupama0503du,stanishev07}, and 
$1.14 \pm 0.01$ mag \citep{misra05,krisciunas07}, respectively.
We must be careful about the time evolution of polarization.
It is thought that the degree of polarization decreases with 
time \citep{wang0301el,wang07,patat09}, but the number of 
multi-epoch spectropolarimetric observations is not sufficient
to clarify the common property of the time evolution.

Keeping this caveat in mind, we can still see the trend that
brighter SNe have smaller line polarization.
As is inferred from spectral models for Type Ia SNe \citep{mazzali07Ia},
brigher SNe have the thinner Si-rich layers, \ie
the degree of the line polarization is smaller in 
SNe with the thinner Si-rich layers.
In addition, SNe with high Si II line velocities tend to
have a high Si II line polarization 
\citep{leonard05,wang06,patat09}.
In high-velocity SNe, the Si-rich layer is thought to be thick
\citep{stehle05,tanaka08Ia}.
We thus suggest that the Si line polarization 
can be used as an indicator of thickness of the Si-rich layers,
or that enough thick Si-rich layers are required to produce 
the sufficient \ion{Si}{ii} line polarization.

SN 2009dc is an extremely luminous SN,
synthesizing $\gsim 1.2 \Msun$ of \Nifs\
($\Delta m_{15}(B)=0.65 \pm 0.03$ mag, \citealt{yamanaka09}).
Despite the large production of \Nifs,
we have detected a moderate degree of line polarization.
The degree of the Si line polarization
($0.5 \% \pm 0.1 \%$, red point in Figure \ref{fig:pdm15})
is higher than expected from the trend seen in other Type Ia SNe
(blue and green points).
This fact suggests that SN 2009dc has enough thick, 
clumpy Si-rich layers above $\gsim 1.2 \Msun$ of \Nifs-rich layers.

It must be cautioned, however, that this discussion relies largely on the 
small number of spectropolarimetry of SN 1991T-like
overluminous Type Ia SN, especially on SN 2001V 
(\citealt{wang07}, see \citealt{matheson08} for spectroscopic behavior).
To firm the conclusion,
more spectropolarimetric observations of SNe
with $\Delta m_{15}(B) = 0.7-1.0$ mag are necessary.

\subsection{Theoretical Models}

We have shown that SN 2009dc shows the small continuum polarization 
and moderate line polarizations.
From the relation between $\Delta m_{15}(B)$ and the \ion{Si}{ii} line 
polarization, it is suggested that SN 2009dc has enough thick, 
clumpy Si-rich layers.
Combined with the low line velocities and the clear detection of 
the strong \ion{C}{ii} line \citep{yamanaka09}, 
we suggest that SN 2009dc has a super-Chandrasekhar mass.

\citet{hillebrandt07} suggested a possibility that
the extremely luminous Type Ia SN 2003fg
is an aspherical explosion of a Chandrasekhar mass white dwarf.
In their scenario, the explosion is triggered at an off-center position,
and nuclear burning preferentially takes place toward one direction.
As a result, the distribution of \Nifs\ is lopsided.
With such a lopsided explosion, 
the extreme luminosity can be explained even with 0.9 $\Msun$ of \Nifs\
when we observe the SN from the \Nifs-rich direction.

Although this explosion model is very aspherical,
the expected continuum polarization may not be very large.
\citet{kasen07} presented theoretical predictions of
the polarization spectra for a detonating failed  
deflagration model by \citet{plewa07}, in which 
the off-center ignition is followed by a surface detonation.
The center of the \Nifs\ distribution is displaced
also in this model. The resultant SN can be brighter when 
it is viewed from the direction in which the detonation takes place.
The predicted continuum polarization is small in every line of sight.
This is understood as following.
In the ellipsoidal geometry, non-zero polarization is expected
because of the difference in the amount of light from the pole and the 
equator (90-degrees away), having orthogonal polarization directions. 
In the off-center model, on the other hand, the difference is mainly 
in 180-degrees away, and thus, the resultant net polarization is small.

It is also worth comparing the \ion{Si}{ii} line polarization 
of this off-center ignition model \citep{plewa07,kasen07}
with the observation ($0.5\% \pm 0.1\%$).
Although the expected polarization depends on the line of sight,
\citet{kasen07} found that the line polarization tends to be high, 
reflecting the aspherical distribution of intermediate mass elements.
In there model, the \ion{Si}{ii} line polarization is
greater than 1 $\%$ in almost half of the possible viewing angles.

Next, we discuss explosion geometry of a super-Chandrasekhar mass white dwarf.
When a super-Chandrasekhar mass is maintained by rapid rotation, 
the rotation speed is limited by the critical rotation at the surface in
the equatorial direction.  
If the rotation is uniform, 
the Chandrasekhar mass is only 1.5 $\Msun$.  
If the white dwarf mass is as large as 2 $\Msun$, 
then the rotation must be differential and the white dwarf prior
to the explosion has an aspherical shape \citep{uenishi03,yoon05}.

The explosion of super-Chandrasekhar mass white dwarf 
was first studied by \citet{steinmetz92}.
They explored pure detonation models and found that 
almost all the materials are burned into Fe-peak elements.
This is incompatible with the observed properties of 
normal and extremely luminous Type Ia SNe.

Recently, \citet{pfannes10a} performed simulations of 
deflagration models of super-Chandrasekhar mass white dwarf.
They found that the deflagration model does not result in a
bright enough SN, and is unlikely to explain the extremely luminous 
Type Ia SNe, as noted by \citet{hillebrandt07}.

Interestingly, \citet{pfannes10b} revised the simulations of 
pure detonation models of super-Chandrasekhar mass white dwarf.
They found that the entire white dwarf
is not burned into Fe-peak elements, and that 
some intermediate mass elements are synthesized.
In their models, the synthesized mass of \Nifs\ is 
sufficiently large ($\sim 1.5 \Msun$) 
to explain the extremely luminous Type Ia SNe.
It is worth testing the spectroscopic and spectropolarimetric 
properties of this model.

Note that a recently proposed scenario involving 
the collision of two white dwarfs could also 
realize super-Chandrasekhar mass ejecta \citep{rosswog09,raskin09}.
However, this scenario predicts a very aspherical explosion,
and thus, it may not be consistent with the small continuum 
polarization level observed in SN 2009dc.

\section{Conclusions}
\label{sec:conclusions}

We have presented the first spectropolarimetric observations of 
a candidate of the super-Chandrasekhar mass Type Ia SN: SN 2009dc.
With the data taken at a later epoch,
ISP is estimated with small uncertainty.
We find that 
the continuum polarization of SN 2009dc is small ($<0.3 \%$).
The degree of line polarization is moderate 
($0.5 \% \pm 0.1 \%$ at \ion{Si}{ii} $\lambda$6355, and  
$0.7 \% \pm 0.1 \%$ at the \ion{Ca}{ii} IR triplet).
These properties of polarization are similar to those of 
normally luminous Type Ia SNe with normal expansion velocities.

The small continuum polarization indicates that 
the explosion is nearly spherically symmetric.
This fact suggests that a very aspherical explosion 
is not a likely scenario for SN 2009dc.

The polarization at the \ion{Si}{ii} and \ion{Ca}{ii} lines
is detected not only at photospheric velocity
($7,200$ \kms) but also at high velocity ($v \gsim 12,000$ \kms).
In addition, polarization data
clearly show a loop in the $Q$-$U$ plane, suggesting non-axisymmetric, 
clumpy distribution of intermediate-mass elements.

Past spectropolarimetric observations of Type Ia SNe 
show the trend that brighter SNe have smaller line polarization
\citep{wang07}.
However, the degree of line polarization of SN 2009dc deviates 
from this trend: it is larger than expected from its extreme brightness.
This fact suggests that there are thick enough,
clumpy Si-rich layers above the thick \Nifs-rich layers 
($\gsim 1.2 \Msun$), and thus,
the progenitor white dwarf has a super-Chandrasekhar mass.

In summary, our spectropolarimetric observations, combined with
photometric and spectroscopic observations \citep{yamanaka09},
suggest that 
SN 2009dc is an explosion of a super-Chandrasekhar mass white dwarf.
The explosion geometry is globally spherically symmetric, 
with clumpy distribution of intermediate-mass elements. 

\acknowledgments
We are grateful to the staff of the Subaru Telescope,
especially Masahiko Hayashi and Hiroshi Terada 
for kindly allowing our observation in ToO time.
We thank the anonymous referee for the helpful suggestions.
M.T. is supported by the JSPS (Japan Society for the Promotion of Science) 
Research Fellowship for Young Scientists.
This research has been supported in part by World Premier
International Research Center Initiative, MEXT,
Japan, and by the Grant-in-Aid for Scientific Research of the JSPS
(18104003, 18540231, 20540226, 20840007) and MEXT (19047004, 20040004).


\begin{thebibliography}{73}
\expandafter\ifx\csname natexlab\endcsname\relax\def\natexlab#1{#1}\fi

\bibitem[{{Akerlof} {et~al.}(2007){Akerlof}, {Miller}, {Peters}, {Thorstensen},
  {Baltay}, {Bauer}, {Rabinowitz}, {Scalzo}, {Rigaudier}, {Pecontal}, {Buton},
  {Copin}, {Gangler}, {Smadja}, {Tao}, {Antilogus}, {Bailey}, {Pain},
  {Pereira}, {Wu}, {Aldering}, {Aragon}, {Bongard}, {Childress}, {Loken},
  {Nugent}, {Perlmutter}, {Runge}, {Thomas}, {Weaver}, {Birchall}, {Cough},
  {Holtzman}, {Rau}, {Kasliwal}, {Gal-Yam}, {Yuan}, \& {Quimby}}]{akerlof07}
{Akerlof}, C., {et~al.} 2007, Central Bureau Electronic Telegrams, 1059, 2

\bibitem[{{Anupama} {et~al.}(2005){Anupama}, {Sahu}, \& {Jose}}]{anupama0503du}
{Anupama}, G.~C., {Sahu}, D.~K., \& {Jose}, J. 2005, \aap, 429, 667

\bibitem[{{Chornock} \& {Filippenko}(2008)}]{chornock08}
{Chornock}, R., \& {Filippenko}, A.~V. 2008, \aj, 136, 2227

\bibitem[{{Gerardy} {et~al.}(2004){Gerardy}, {H{\"o}flich}, {Fesen}, {Marion},
  {Nomoto}, {Quimby}, {Schaefer}, {Wang}, \& {Wheeler}}]{gerardy04}
{Gerardy}, C.~L., {et~al.} 2004, \apj, 607, 391

\bibitem[{{Harutyunyan} {et~al.}(2009){Harutyunyan}, {Elias-Rosa}, \&
  {Benetti}}]{harutyunyan09}
{Harutyunyan}, A., {Elias-Rosa}, N., \& {Benetti}, S. 2009, Central Bureau
  Electronic Telegrams, 1768, 1

\bibitem[{{Hicken} {et~al.}(2007){Hicken}, {Garnavich}, {Prieto}, {Blondin},
  {DePoy}, {Kirshner}, \& {Parrent}}]{hicken07}
{Hicken}, M., {Garnavich}, P.~M., {Prieto}, J.~L., {Blondin}, S., {DePoy},
  D.~L., {Kirshner}, R.~P., \& {Parrent}, J. 2007, \apjl, 669, L17

\bibitem[{{Hillebrandt} \& {Niemeyer}(2000)}]{hillebrandt00}
{Hillebrandt}, W., \& {Niemeyer}, J.~C. 2000, \araa, 38, 191

\bibitem[{{Hillebrandt} {et~al.}(2007){Hillebrandt}, {Sim}, \&
  {R{\"o}pke}}]{hillebrandt07}
{Hillebrandt}, W., {Sim}, S.~A., \& {R{\"o}pke}, F.~K. 2007, \aap, 465, L17

\bibitem[{{H{\"o}flich}(1991)}]{hoeflich91}
{H{\"o}flich}, P. 1991, \aap, 246, 481

\bibitem[{{Howell} {et~al.}(2001){Howell}, {H{\"o}flich}, {Wang}, \&
  {Wheeler}}]{howell01}
{Howell}, D.~A., {H{\"o}flich}, P., {Wang}, L., \& {Wheeler}, J.~C. 2001, \apj,
  556, 302

\bibitem[{{Howell} {et~al.}(2006){Howell}, {Sullivan}, {Nugent}, {Ellis},
  {Conley}, {Le Borgne}, {Carlberg}, {Guy}, {Balam}, {Basa}, {Fouchez}, {Hook},
  {Hsiao}, {Neill}, {Pain}, {Perrett}, \& {Pritchet}}]{howell06}
{Howell}, D.~A., {et~al.} 2006, \nat, 443, 308

\bibitem[{{Jeffrey}(1989)}]{jeffery89}
{Jeffrey}, D.~J. 1989, \apjs, 71, 951

\bibitem[{{Jha} {et~al.}(2006){Jha}, {Kirshner}, {Challis}, {Garnavich},
  {Matheson}, {Soderberg}, {Graves}, {Hicken}, {Alves}, {Arce}, {Balog},
  {Barmby}, {Barton}, {Berlind}, {Bragg}, {Brice{\~n}o}, {Brown}, {Buckley},
  {Caldwell}, {Calkins}, {Carter}, {Concannon}, {Donnelly}, {Eriksen},
  {Fabricant}, {Falco}, {Fiore}, {Garcia}, {G{\'o}mez}, {Grogin}, {Groner},
  {Groot}, {Haisch}, {Hartmann}, {Hergenrother}, {Holman}, {Huchra},
  {Jayawardhana}, {Jerius}, {Kannappan}, {Kim}, {Kleyna}, {Kochanek},
  {Koranyi}, {Krockenberger}, {Lada}, {Luhman}, {Luu}, {Macri}, {Mader},
  {Mahdavi}, {Marengo}, {Marsden}, {McLeod}, {McNamara}, {Megeath}, {Moraru},
  {Mossman}, {Muench}, {Mu{\~n}oz}, {Muzerolle}, {Naranjo}, {Nelson-Patel},
  {Pahre}, {Patten}, {Peters}, {Peters}, {Raymond}, {Rines}, {Schild},
  {Sobczak}, {Spahr}, {Stauffer}, {Stefanik}, {Szentgyorgyi}, {Tollestrup},
  {V{\"a}is{\"a}nen}, {Vikhlinin}, {Wang}, {Willner}, {Wolk}, {Zajac}, {Zhao},
  \& {Stanek}}]{jha06}
{Jha}, S., {et~al.} 2006, \aj, 131, 527

\bibitem[{{Kasen} {et~al.}(2003){Kasen}, {Nugent}, {Wang}, {Howell}, {Wheeler},
  {H{\"o}flich}, {Baade}, {Baron}, \& {Hauschildt}}]{kasen03}
{Kasen}, D., {et~al.} 2003, \apj, 593, 788

\bibitem[{{Kasen} \& {Plewa}(2007)}]{kasen07}
{Kasen}, D., \& {Plewa}, T. 2007, \apj, 662, 459

\bibitem[{{Kashikawa} {et~al.}(2002){Kashikawa}, {Aoki}, {Asai}, {Ebizuka},
  {Inata}, {Iye}, {Kawabata}, {Kosugi}, {Ohyama}, {Okita}, {Ozawa}, {Saito},
  {Sasaki}, {Sekiguchi}, {Shimizu}, {Taguchi}, {Takata}, {Yadoumaru}, \&
  {Yoshida}}]{kashikawa02}
{Kashikawa}, N., {et~al.} 2002, \pasj, 54, 819

\bibitem[{{Kawabata} {et~al.}(2002){Kawabata}, {Jeffery}, {Iye}, {Ohyama},
  {Kosugi}, {Kashikawa}, {Ebizuka}, {Sasaki}, {Sekiguchi}, {Nomoto}, {Mazzali},
  {Deng}, {Maeda}, {Umeda}, {Aoki}, {Saito}, {Takata}, {Yoshida}, {Asai},
  {Inata}, {Okita}, {Ota}, {Ozawa}, {Shimizu}, {Taguchi}, {Yadoumaru},
  {Misawa}, {Nakata}, {Yamada}, {Tanaka}, \& {Kodama}}]{kawabata02}
{Kawabata}, K.~S., {et~al.} 2002, \apjl, 580, L39

\bibitem[{{Krisciunas} {et~al.}(2007){Krisciunas}, {Garnavich}, {Stanishev},
  {Suntzeff}, {Prieto}, {Espinoza}, {Gonzalez}, {Salvo}, {Elias de la Rosa},
  {Smartt}, {Maund}, \& {Kudritzki}}]{krisciunas07}
{Krisciunas}, K., {et~al.} 2007, \aj, 133, 58

\bibitem[{{Landi degl'Innocenti}(2002)}]{landi02}
{Landi degl'Innocenti}, E. 2002, in Astrophysical Spectropolarimetry, ed.
  J.~{Trujillo-Bueno}, F.~{Moreno-Insertis}, \& F.~{S{\'a}nchez}, 1--53

\bibitem[{{Leonard} {et~al.}(2005){Leonard}, {Li}, {Filippenko}, {Foley}, \&
  {Chornock}}]{leonard05}
{Leonard}, D.~C., {Li}, W., {Filippenko}, A.~V., {Foley}, R.~J., \& {Chornock},
  R. 2005, \apj, 632, 450

\bibitem[{{Maeda} \& {Iwamoto}(2009)}]{maedaiwamoto09}
{Maeda}, K., \& {Iwamoto}, K. 2009, \mnras, 394, 239

\bibitem[{{Maeda} {et~al.}(2009){Maeda}, {Kawabata}, {Li}, {Tanaka}, {Mazzali},
  {Hattori}, {Nomoto}, \& {Filippenko}}]{maeda09}
{Maeda}, K., {Kawabata}, K., {Li}, W., {Tanaka}, M., {Mazzali}, P.~A.,
  {Hattori}, T., {Nomoto}, K., \& {Filippenko}, A.~V. 2009, \apj, 690, 1745

\bibitem[{{Marion} {et~al.}(2006){Marion}, {H{\"o}flich}, {Wheeler},
  {Robinson}, {Gerardy}, \& {Vacca}}]{marion06}
{Marion}, G.~H., {H{\"o}flich}, P., {Wheeler}, J.~C., {Robinson}, E.~L.,
  {Gerardy}, C.~L., \& {Vacca}, W.~D. 2006, \apj, 645, 1392

\bibitem[{{Marion} {et~al.}(2009){Marion}, {Garnavich}, {Challis}, {Calkins},
  \& {Peters}}]{marion09}
{Marion}, H., {Garnavich}, P., {Challis}, P., {Calkins}, M., \& {Peters}, W.
  2009, Central Bureau Electronic Telegrams, 1776, 1

\bibitem[{{Matheson} {et~al.}(2008){Matheson}, {Kirshner}, {Challis}, {Jha},
  {Garnavich}, {Berlind}, {Calkins}, {Blondin}, {Balog}, {Bragg}, {Caldwell},
  {Dendy Concannon}, {Falco}, {Graves}, {Huchra}, {Kuraszkiewicz}, {Mader},
  {Mahdavi}, {Phelps}, {Rines}, {Song}, \& {Wilkes}}]{matheson08}
{Matheson}, T., {et~al.} 2008, \aj, 135, 1598

\bibitem[{{Maund} {et~al.}(2007){Maund}, {Wheeler}, {Patat}, {Baade}, {Wang},
  \& {H{\"o}flich}}]{maund0705bf}
{Maund}, J.~R., {Wheeler}, J.~C., {Patat}, F., {Baade}, D., {Wang}, L., \&
  {H{\"o}flich}, P. 2007, \mnras, 381, 201

\bibitem[{{Mazzali}(2001)}]{mazzali0190N}
{Mazzali}, P.~A. 2001, \mnras, 321, 341

\bibitem[{{Mazzali} {et~al.}(2007){Mazzali}, {R{\"o}pke}, {Benetti}, \&
  {Hillebrandt}}]{mazzali07Ia}
{Mazzali}, P.~A., {R{\"o}pke}, F.~K., {Benetti}, S., \& {Hillebrandt}, W. 2007,
  Science, 315, 825

\bibitem[{{McCall}(1984)}]{mccall84}
{McCall}, M.~L. 1984, \mnras, 210, 829

\bibitem[{{Misra} {et~al.}(2005){Misra}, {Kamble}, {Bhattacharya}, \&
  {Sagar}}]{misra05}
{Misra}, K., {Kamble}, A.~P., {Bhattacharya}, D., \& {Sagar}, R. 2005, \mnras,
  360, 662

\bibitem[{{Nomoto} {et~al.}(1994){Nomoto}, {Yamaoka}, {Shigeyama}, {Kumagai},
  \& {Tsujimoto}}]{nomoto94Ia}
{Nomoto}, K., {Yamaoka}, H., {Shigeyama}, T., {Kumagai}, S., \& {Tsujimoto}, T.
  1994, in Supernovae, ed. S.~A. {Bludman}, R.~{Mochkovitch}, \&
  J.~{Zinn-Justin}, 199--+

\bibitem[{{Oke}(1990)}]{oke90}
{Oke}, J.~B. 1990, \aj, 99, 1621

\bibitem[{{Patat} {et~al.}(2009){Patat}, {Baade}, {H{\"o}flich}, {Maund},
  {Wang}, \& {Wheeler}}]{patat09}
{Patat}, F., {Baade}, D., {H{\"o}flich}, P., {Maund}, J.~R., {Wang}, L., \&
  {Wheeler}, J.~C. 2009, \aap, 508, 229

\bibitem[{{Patat} \& {Romaniello}(2006)}]{patat06}
{Patat}, F., \& {Romaniello}, M. 2006, \pasp, 118, 146

\bibitem[{{Perlmutter} {et~al.}(1999){Perlmutter}, {Aldering}, {Goldhaber},
  {Knop}, {Nugent}, {Castro}, {Deustua}, {Fabbro}, {Goobar}, {Groom}, {Hook},
  {Kim}, {Kim}, {Lee}, {Nunes}, {Pain}, {Pennypacker}, {Quimby}, {Lidman},
  {Ellis}, {Irwin}, {McMahon}, {Ruiz-Lapuente}, {Walton}, {Schaefer}, {Boyle},
  {Filippenko}, {Matheson}, {Fruchter}, {Panagia}, {Newberg}, {Couch}, \& {The
  Supernova Cosmology Project}}]{perlmutter99}
{Perlmutter}, S., {et~al.} 1999, \apj, 517, 565

\bibitem[{{Pfannes} {et~al.}(2010{\natexlab{a}}){Pfannes}, {Niemeyer}, \&
  {Schmidt}}]{pfannes10b}
{Pfannes}, J.~M.~M., {Niemeyer}, J.~C., \& {Schmidt}, W. 2010{\natexlab{a}},
  \aap, 509, A75+

\bibitem[{{Pfannes} {et~al.}(2010{\natexlab{b}}){Pfannes}, {Niemeyer},
  {Schmidt}, \& {Klingenberg}}]{pfannes10a}
{Pfannes}, J.~M.~M., {Niemeyer}, J.~C., {Schmidt}, W., \& {Klingenberg}, C.
  2010{\natexlab{b}}, \aap, 509, A74+

\bibitem[{{Phillips}(1993)}]{phillips93}
{Phillips}, M.~M. 1993, \apjl, 413, L105

\bibitem[{{Phillips} {et~al.}(1999){Phillips}, {Lira}, {Suntzeff}, {Schommer},
  {Hamuy}, \& {Maza}}]{phillips99}
{Phillips}, M.~M., {Lira}, P., {Suntzeff}, N.~B., {Schommer}, R.~A., {Hamuy},
  M., \& {Maza}, J. 1999, \aj, 118, 1766

\bibitem[{{Plewa}(2007)}]{plewa07}
{Plewa}, T. 2007, \apj, 657, 942

\bibitem[{{Puckett} {et~al.}(2009){Puckett}, {Moore}, {Newton}, \&
  {Orff}}]{puckett09}
{Puckett}, T., {Moore}, R., {Newton}, J., \& {Orff}, T. 2009, Central Bureau
  Electronic Telegrams, 1762, 1

\bibitem[{{Raskin} {et~al.}(2009){Raskin}, {Timmes}, {Scannapieco}, {Diehl}, \&
  {Fryer}}]{raskin09}
{Raskin}, C., {Timmes}, F.~X., {Scannapieco}, E., {Diehl}, S., \& {Fryer}, C.
  2009, \mnras, 399, L156

\bibitem[{{Rau} {et~al.}(2007){Rau}, {Kasliwal}, \& {Gal-Yam}}]{rau07}
{Rau}, A., {Kasliwal}, M., \& {Gal-Yam}, A. 2007, Central Bureau Electronic
  Telegrams, 1059, 3

\bibitem[{{Riess} {et~al.}(1998){Riess}, {Filippenko}, {Challis},
  {Clocchiatti}, {Diercks}, {Garnavich}, {Gilliland}, {Hogan}, {Jha},
  {Kirshner}, {Leibundgut}, {Phillips}, {Reiss}, {Schmidt}, {Schommer},
  {Smith}, {Spyromilio}, {Stubbs}, {Suntzeff}, \& {Tonry}}]{riess98}
{Riess}, A.~G., {et~al.} 1998, \aj, 116, 1009

\bibitem[{{Rosswog} {et~al.}(2009){Rosswog}, {Kasen}, {Guillochon}, \&
  {Ramirez-Ruiz}}]{rosswog09}
{Rosswog}, S., {Kasen}, D., {Guillochon}, J., \& {Ramirez-Ruiz}, E. 2009,
  \apjl, 705, L128

\bibitem[{{Scalzo} {et~al.}(2010){Scalzo}, {Aldering}, {Antilogus}, {Aragon},
  {Bailey}, {Bongard}, {Buton}, {Childress}, {Chotard}, {Copin}, {Fakhouri},
  {Gal-Yam}, {Gangler}, {Hoyer}, {Kasliwal}, {Loken}, {Nugent}, {Pain},
  {Pecontal}, {Pereira}, {Perlmutter}, {Rabinowitz}, {Rau}, {Rigaudier},
  {Runge}, {Smadja}, {Tao}, {Thomas}, {Weaver}, \& {Wu}}]{scalzo10}
{Scalzo}, R.~A., {et~al.} 2010, ApJ, in press (arXiv:1003.2217)

\bibitem[{{Schlegel} {et~al.}(1998){Schlegel}, {Finkbeiner}, \&
  {Davis}}]{schlegel98}
{Schlegel}, D.~J., {Finkbeiner}, D.~P., \& {Davis}, M. 1998, \apj, 500, 525

\bibitem[{{Schmidt} {et~al.}(1992){Schmidt}, {Elston}, \&
  {Lupie}}]{schmidt92pol}
{Schmidt}, G.~D., {Elston}, R., \& {Lupie}, O.~L. 1992, \aj, 104, 1563

\bibitem[{{Serkowski} {et~al.}(1975){Serkowski}, {Mathewson}, \&
  {Ford}}]{serkowski75}
{Serkowski}, K., {Mathewson}, D.~L., \& {Ford}, V.~L. 1975, \apj, 196, 261

\bibitem[{{Shapiro} \& {Sutherland}(1982)}]{shapiro82}
{Shapiro}, P.~R., \& {Sutherland}, P.~G. 1982, \apj, 263, 902

\bibitem[{{Silverman} {et~al.}(2010){Silverman}, {Ganeshalingam}, {Li},
  {Filippenko}, {Miller}, \& {Poznanski}}]{silverman10}
{Silverman}, J.~M., {Ganeshalingam}, M., {Li}, W., {Filippenko}, A.~V.,
  {Miller}, A.~A., \& {Poznanski}, D. 2010, submitted to MNRAS
  (arXiv:1003.2417)

\bibitem[{{Stanishev} {et~al.}(2007){Stanishev}, {Goobar}, {Benetti}, {Kotak},
  {Pignata}, {Navasardyan}, {Mazzali}, {Amanullah}, {Garavini}, {Nobili},
  {Qiu}, {Elias-Rosa}, {Ruiz-Lapuente}, {Mendez}, {Meikle}, {Patat},
  {Pastorello}, {Altavilla}, {Gustafsson}, {Harutyunyan}, {Iijima},
  {Jakobsson}, {Kichizhieva}, {Lundqvist}, {Mattila}, {Melinder}, {Pavlenko},
  {Pavlyuk}, {Sollerman}, {Tsvetkov}, {Turatto}, \&
  {Hillebrandt}}]{stanishev07}
{Stanishev}, V., {et~al.} 2007, \aap, 469, 645

\bibitem[{{Stehle} {et~al.}(2005){Stehle}, {Mazzali}, {Benetti}, \&
  {Hillebrandt}}]{stehle05}
{Stehle}, M., {Mazzali}, P.~A., {Benetti}, S., \& {Hillebrandt}, W. 2005,
  \mnras, 360, 1231

\bibitem[{{Steinmetz} {et~al.}(1992){Steinmetz}, {Muller}, \&
  {Hillebrandt}}]{steinmetz92}
{Steinmetz}, M., {Muller}, E., \& {Hillebrandt}, W. 1992, \aap, 254, 177

\bibitem[{{Tanaka} {et~al.}(2008{\natexlab{a}}){Tanaka}, {Kawabata}, {Maeda},
  {Hattori}, \& {Nomoto}}]{tanaka0807gr}
{Tanaka}, M., {Kawabata}, K.~S., {Maeda}, K., {Hattori}, T., \& {Nomoto}, K.
  2008{\natexlab{a}}, \apj, 689, 1191

\bibitem[{{Tanaka} {et~al.}(2009){Tanaka}, {Kawabata}, {Maeda}, {Iye},
  {Hattori}, {Pian}, {Nomoto}, {Mazzali}, \& {Tominaga}}]{tanaka0905bf}
{Tanaka}, M., {et~al.} 2009, \apj, 699, 1119

\bibitem[{{Tanaka} {et~al.}(2008{\natexlab{b}}){Tanaka}, {Mazzali}, {Benetti},
  {Nomoto}, {Elias-Rosa}, {Kotak}, {Pignata}, {Stanishev}, \&
  {Hachinger}}]{tanaka08Ia}
---. 2008{\natexlab{b}}, \apj, 677, 448

\bibitem[{{Thomas} {et~al.}(2007){Thomas}, {Aldering}, {Antilogus}, {Aragon},
  {Bailey}, {Baltay}, {Baron}, {Bauer}, {Buton}, {Bongard}, {Copin}, {Gangler},
  {Gilles}, {Kessler}, {Loken}, {Nugent}, {Pain}, {Parrent}, {P{\'e}contal},
  {Pereira}, {Perlmutter}, {Rabinowitz}, {Rigaudier}, {Runge}, {Scalzo},
  {Smadja}, {Wang}, \& {Weaver}}]{thomas07}
{Thomas}, R.~C., {et~al.} 2007, \apjl, 654, L53

\bibitem[{{Tinbergen}(1996)}]{tinbergen96}
{Tinbergen}, J. 1996, {Astronomical Polarimetry} (Cambridge, UK: Cambridge
  University Press)

\bibitem[{{Turatto} {et~al.}(2003){Turatto}, {Benetti}, \&
  {Cappellaro}}]{turatto03}
{Turatto}, M., {Benetti}, S., \& {Cappellaro}, E. 2003, in From Twilight to
  Highlight: The Physics of Supernovae, ed. W.~{Hillebrandt} \&
  B.~{Leibundgut}, 200--+

\bibitem[{{Turnshek} {et~al.}(1990){Turnshek}, {Bohlin}, {Williamson}, {Lupie},
  {Koornneef}, \& {Morgan}}]{turnshek90}
{Turnshek}, D.~A., {Bohlin}, R.~C., {Williamson}, II, R.~L., {Lupie}, O.~L.,
  {Koornneef}, J., \& {Morgan}, D.~H. 1990, \aj, 99, 1243

\bibitem[{{Uenishi} {et~al.}(2003){Uenishi}, {Nomoto}, \&
  {Hachisu}}]{uenishi03}
{Uenishi}, T., {Nomoto}, K., \& {Hachisu}, I. 2003, \apj, 595, 1094

\bibitem[{{Wang} {et~al.}(2003{\natexlab{a}}){Wang}, {Baade}, {H{\"o}flich},
  {Khokhlov}, {Wheeler}, {Kasen}, {Nugent}, {Perlmutter}, {Fransson}, \&
  {Lundqvist}}]{wang0301el}
{Wang}, L., {et~al.} 2003{\natexlab{a}}, \apj, 591, 1110

\bibitem[{{Wang} {et~al.}(2003{\natexlab{b}}){Wang}, {Baade}, {H{\"o}flich}, \&
  {Wheeler}}]{wang0302ap}
{Wang}, L., {Baade}, D., {H{\"o}flich}, P., \& {Wheeler}, J.~C.
  2003{\natexlab{b}}, \apj, 592, 457

\bibitem[{{Wang} {et~al.}(2006){Wang}, {Baade}, {H{\"o}flich}, {Wheeler},
  {Kawabata}, {Khokhlov}, {Nomoto}, \& {Patat}}]{wang06}
{Wang}, L., {Baade}, D., {H{\"o}flich}, P., {Wheeler}, J.~C., {Kawabata}, K.,
  {Khokhlov}, A., {Nomoto}, K., \& {Patat}, F. 2006, \apj, 653, 490

\bibitem[{{Wang} {et~al.}(2007){Wang}, {Baade}, \& {Patat}}]{wang07}
{Wang}, L., {Baade}, D., \& {Patat}, F. 2007, Science, 315, 212

\bibitem[{{Wang} {et~al.}(2001){Wang}, {Howell}, {H{\"o}flich}, \&
  {Wheeler}}]{wang01}
{Wang}, L., {Howell}, D.~A., {H{\"o}flich}, P., \& {Wheeler}, J.~C. 2001, \apj,
  550, 1030

\bibitem[{{Wang} \& {Wheeler}(2008)}]{wang08}
{Wang}, L., \& {Wheeler}, J.~C. 2008, \araa, 46, 433

\bibitem[{{Wang} {et~al.}(1997){Wang}, {Wheeler}, \& {Hoeflich}}]{wang97}
{Wang}, L., {Wheeler}, J.~C., \& {Hoeflich}, P. 1997, \apjl, 476, L27+

\bibitem[{{Whittet} {et~al.}(1992){Whittet}, {Martin}, {Hough}, {Rouse},
  {Bailey}, \& {Axon}}]{whittet92}
{Whittet}, D.~C.~B., {Martin}, P.~G., {Hough}, J.~H., {Rouse}, M.~F., {Bailey},
  J.~A., \& {Axon}, D.~J. 1992, \apj, 386, 562

\bibitem[{{Yamanaka} {et~al.}(2009){Yamanaka}, {Kawabata}, {Kinugasa},
  {Tanaka}, {Imada}, {Maeda}, {Nomoto}, {Arai}, {Chiyonobu}, {Fukazawa},
  {Hashimoto}, {Honda}, {Ikejiri}, {Itoh}, {Kamata}, {Kawai}, {Komatsu},
  {Konishi}, {Kuroda}, {Miyamoto}, {Miyazaki}, {Nagae}, {Nakaya}, {Ohsugi},
  {Omodaka}, {Sakai}, {Sasada}, {Suzuki}, {Taguchi}, {Takahashi}, {Tanaka},
  {Uemura}, {Yamashita}, {Yanagisawa}, \& {Yoshida}}]{yamanaka09}
{Yamanaka}, M., {et~al.} 2009, \apjl, 707, L118

\bibitem[{{Yoon} \& {Langer}(2005)}]{yoon05}
{Yoon}, S.-C., \& {Langer}, N. 2005, \aap, 435, 967

\bibitem[{{Yuan} {et~al.}(2007){Yuan}, {Quimby}, {Peters}, \&
  {Thorstensen}}]{yuan07}
{Yuan}, F., {Quimby}, R., {Peters}, C., \& {Thorstensen}, J. 2007, Central
  Bureau Electronic Telegrams, 1059, 1

\end{thebibliography}
\end{document}